\documentclass[12pt]{iopart}
\usepackage{iopams}
\usepackage{graphicx}
\begin{document}

\title{Fractional diffusion in periodic potentials}

\author{E Heinsalu$^{1, 2}$, M Patriarca$^1$, I Goychuk$^1$ and P H\"anggi$^1$}

\address{$^1$ Institut f\"ur Physik, Universit\"at Augsburg, Universit\"atsstr. 1, D-86135 Augsburg, Germany}
\address{$^2$ Institute of Theoretical Physics, Tartu University, 4 T\"ahe Street, 51010 Tartu, Estonia}


\begin{abstract}
Fractional, anomalous diffusion in space-periodic potentials is investigated. The
analytical solution for the effective, fractional diffusion
coefficient in an arbitrary periodic potential is obtained in
closed form in terms of two quadratures. This theoretical result is corroborated by numerical
simulations for different shapes of the periodic potential. Normal
and fractional spreading processes are contrasted via their  time
evolution of the corresponding probability densities in state
space. While there are distinct differences occurring at small evolution times,
a re-scaling of time yields a mutual matching between the long-time behaviors
of normal and fractional diffusion.
\end{abstract}






\section{Introduction} \label{introduction}

In 1905 Pearson proposed what we now know as a random walk
\cite{pearson1905}. For a one-dimensional system the problem can
be formulated in the following way: A particle jumps at each point
of time from its current position $x$ to the position $x + \Delta
x$ with probability $p$, or  $x - \Delta x$ with probability
$1-p$. The approach towards diffusion theory, pioneered by
Einstein, relies on postulates very similar to the ones for the
random walk, leading to the same results when the jump width
$\Delta x \to 0$ \cite{einstein1905}. However, in many situations,
the assumptions used by Einstein and Pearson do not hold; one of
such examples is the transport of charge carriers in amorphous
semiconductors when exposed to an electric field.

Sixty years after Pearson, in 1965, Montroll and Weiss introduced
the theory of continuous time random walks (CTRW)
\cite{MontrollWeiss}. It was applied to the transport in
semiconductors in works by Scher and Lax \cite{ScherLax}, and
Scher and Montroll \cite{ScherMontroll}. Due to its historical
importance and vivid clarity we recall here the definition of the
CTRW as given by Scher and Montroll \cite{ScherMontroll}: \textit{In our model we postulate our
material to be divided into a regular lattice of equivalent cells,
with each cell containing many randomly distributed localized
sites available for hopping carriers. Carrier transport is a
succession of carrier hops from one localized site to another and
finally from one cell to another. We define the hopping time to be
the time interval between the moment of arrival of a carrier into
one cell and the moment of arrival into the next cell into which
it lands. The random distribution of sites and hence the disorder
of an amorphous material is incorporated into a hopping-time
distribution function $\psi (\tau)$}. The appropriate distribution
$\psi (\tau)$, leading to the agreement with the experiments, was
shown to possess the power-law form: $\psi (\tau) \propto \tau^{-1
-\alpha}$ with $\alpha \in (0, 1)$ \cite{ScherMontroll,
shlesinger1974, scher1991}. For this range of the fractional
exponent $\alpha$ all the moments of the distribution $\psi
(\tau)$ diverge and the corresponding process has no
characteristic time scale, thus exhibiting the phenomenon of {\it aging}.
As a result, the process undergoes
subdiffusion \cite{bouchaud1990R, metzler2000R, sokolov2002},
i.e., the mean square displacement grows in the absence of an
external force slower than linearly in time, $\langle \delta r^2
(t) \rangle \sim t^\alpha$ ($0 < \alpha < 1$).

In the original study of the fractional transport in the context
of anomalous transport in semiconductors an ensemble of carriers executing a random walk,
when biased by an electric field, was studied \cite{ScherMontroll,
ScherLax}. In the present paper our focus is different: We instead address the problem of the
carriers executing the random walk in a spatially varying periodic potential. This  situation
is representative  for  many applications occurring in areas such as in condensed matter physics, chemical
physics, nanotechnology, and molecular biology, to name but a few. For those applications it is of utmost
importance to account for the spatial variation of the transport process
\cite{100years,HAM,HM,freund,sokolov,HTB}.

Our work is is set up as follows:
In section~\ref{sec-model} we propose the model and define the
theoretical and numerical problem. In section~\ref{sec-CTRW} we
recall some prior results  about the biased CTRW \cite{metzler1998}
and a CTRW proceeding in a washboard potential \cite{letter}. In
section~\ref{sec-result} the formula for the effective fractional
diffusion coefficient in a periodic potential $U_0(x)=U_0(x+L)$
with period $L$ is derived and the theoretical result is
corroborated by numerical simulations of the CTRW for different
shapes of periodic potentials. Finally, we address the problem of
particles spreading anomalously in a periodic potential also in
the light of the time evolution of the space probability density, as
compared to the case with  normal diffusion.


\section{Set up of the model} \label{sec-model}

\begin{figure}[b]
\centering
\includegraphics[width=10.0cm]{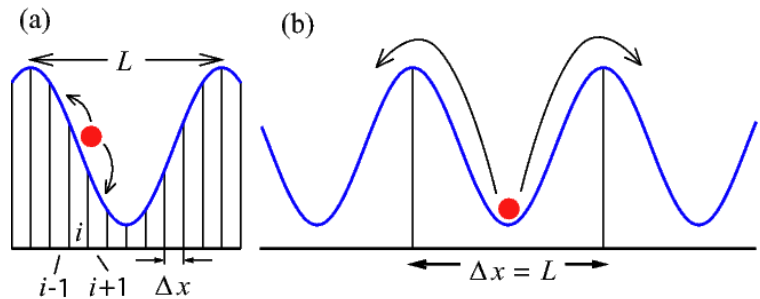}
\caption{(color online) CTRW in a periodic cosine-potential. Two different
possibilities to introduce the one-dimensional lattice: (a) The
lattice period $\Delta x$ is much smaller than the potential
period $L$. The particle at site $i$ hops to site $i+1$, or $i-1$, respectively.
(b)The lattice step size $\Delta x$ is equal to the potential period $L$
and the lattice sites are centered at the potential minima.
The particle performs hops from one potential minima to one of the
neighboring ones.} \label{model}
\end{figure}

Following the general picture of the CTRW we introduce a
one-dimensional lattice $\{ x_i = i \Delta x\}$ with a lattice
period $\Delta x $ and $i = 0, \pm 1, \pm 2, \ldots$ After a
random residence time $\tau$ a particle at site $i$ hops to site
$i \pm 1$ with a probability $q_i^\pm$ (see
figure~\ref{model}(a)): The sites here correspond to the cells in
reference~\cite{ScherMontroll}. The random time $\tau$ is
extracted from a residence time distribution $\psi (\tau)$. A
suitable possible choice for $\psi (\tau)$ is a Mittag-Leffler
distribution defined by
\begin{equation} \label{ML}
\psi_i(\tau) = -\frac{\mathrm{d}}{\mathrm{d} \tau} E_{\alpha}(
-(\nu_i \tau)^{\alpha}) \, , \quad \mathrm{with} \quad
E_{\alpha}(-(\nu_i \tau)^{\alpha}) = \sum_{n = 0}^{\infty}
\frac{[-(\nu_i \tau)^{\alpha}]^n}{ \Gamma(n \alpha + 1)} \, .
\end{equation}
The quantity $\nu _i$ is the time-scaling parameter at lattice site $i$.

The CTRW with the Mittag-Leffler residence time density can be
described through a fractional master equation for the site
populations $P_i(t)$ \cite{letter, HilferAnton}; i.e.,
\begin{equation} \label{FME2}
D_{*}^{\alpha} P_i(t) = f_{i-1} \, P_{i-1}(t) + g_{i+1} \,
P_{i+1}(t) - (f_i + g_i) \, P_i(t) \, ,
\end{equation}
with the Caputo fractional derivative  $D_{*}^{\alpha}$ \cite{GorenfloMainardi} on the
left-hand side  defined by
\begin{equation}
D_{*}^{\alpha} \chi (t) = \frac{1}{\Gamma(1 - \alpha)} \int_0^t
\mathrm{d} t' \frac{1}{(t - t')^\alpha} \frac{\partial}{\partial
t'} \chi (t') \, .
\end{equation}
The quantities $f_i = q_i^{+} \nu_i^\alpha$ and $g_i = q_i^{-}
\nu_i^\alpha$ in the fractional master equation (\ref{FME2}) are
referred to as the fractional forward and backward rates. Using
the normalization condition for the splitting probabilities, i.e.,
$q_i^+ + q_i^- = 1$, one obtains that
\begin{eqnarray}\label{q-nu}
q_i^+ &=& f_i/(f_i + g_i) \, , \qquad  q_i^- = g_i/(f_i + g_i)  \,
, \\
\nu_i &=& (f_i + g_i)^{1/\alpha} \, .
\end{eqnarray}
For an arbitrary shaped potential landscape $U(x)$ the fractional rates can be
chosen as
\begin{eqnarray} \label{f-rate1}
f_i &= (\kappa_{\alpha}/\Delta x ^2) \exp [- \beta(U_{i + 1} -
U_i)/2] \, ,  \\
g_i &= (\kappa_{\alpha}/\Delta x ^2) \exp [- \beta (U_{i - 1} -
U_i)/2] \, . \label{f-rate2}
\end{eqnarray}
Here $U_i \equiv U(i \Delta x)$ and $\beta = 1/k_\mathrm{B} T$ is
the inverse temperature; $\kappa_{\alpha}$ is the fractional free
diffusion coefficient with dimension $\mathrm{cm}^2
\mathrm{s}^{-\alpha}$. The form (\ref{f-rate1})-(\ref{f-rate2}) of
the fractional rates ensures that the Boltzmann detailed balance relation is
satisfied, i.e., $f_{i-1}/g_i= \exp[\beta(U_{i-1}-U_i)]$. The
lattice period $\Delta x$ must fulfill the condition $U''(x) \Delta
x \ll 2 U'(x)$ in order to ensure the smoothness of the potential.
In the case of a periodic potential this implies in particular
that the lattice step size $\Delta x$ is much smaller than the
potential period $L$, $\Delta x \ll L$. Furthermore, in order to
recover the continuous limit addressed below, the condition $|\beta (U_{i\pm 1} -
U_i)|\ll 1$ must be obeyed.

In the space-continuous limit the CTRW with the Mittag-Leffler residence
time density can be described through the fractional Fokker-Planck
equation \cite{metzler2000R, letter, metzler1999, barkai2001},
\begin{eqnarray}\label{FFPE}
D_{*}^{\alpha} P(x,t)= \left [ \frac{\partial}{\partial x}
\frac{U'(x)}{\eta_\alpha} + \kappa _\alpha \frac{\partial
^2}{\partial x^2} \right ] P(x, t) \, .
\end{eqnarray}
Here, $P(x,t)$ is the probability density and a prime stands for
the derivative with respect to the space coordinate. The quantity
$\eta_\alpha$ denotes the generalized friction coefficient,
possessing the dimension $\mathrm{kg} \, \mathrm{s}^{\alpha -2}$.
It is related to the bare fractional anomalous diffusion coefficient $\kappa _\alpha$
through $\eta_\alpha \kappa_\alpha = k_\mathrm{B} T$.

With this material at hand we have defined our theoretical problem as well the
numerical procedure. In the simulations of the CTRW we use  for
$\alpha \in (0, 0.8]$ a Pareto residence time distribution, i.e.,
\begin{equation} \label{pareto}
\psi_i(\tau) = -\frac{\mathrm{d}}{\mathrm{d}\tau} P_\alpha (\nu_i
\tau) \, , \quad \mathrm{with} \quad  P_\alpha (\nu_i \tau) =
\frac{1}{\left[1+\Gamma(1-\alpha)^{1/\alpha}\nu_i
\tau\right]^\alpha } \, ,
\end{equation}
instead of the Mittag-Leffler one, as for every $0< \alpha <1$ the
long time behavior of the system is determined solely by the tail
of the residence time distribution \cite{MainardiFNL}. For $\alpha
> 0.8$ the Mittag-Leffler density~(\ref{ML}) is employed. For $\alpha = 1$
the latter one transforms into the exponential distribution,
covering the regime of normal overdamped Brownian motion. The
spatial lattice step in our simulations is $\Delta x = 0.001$,
measured in units of the spatial period $L$. The energy is
measured in units of the potential amplitude $A$, and the time
unit is set as $\tau_0 = (\eta_\alpha L^2 /A)^{1/\alpha}$. For a
detailed description of the algorithm for the numerical
simulations and of the employment of the Pareto or Mittag-Leffler
distribution, we refer the readers to the comprehensive work in
reference~\cite{heinsalu2006b}.


\section{Biased CTRW  and CTRW in a washboard potential} \label{sec-CTRW}

\subsection{Biased CTRW} \label{subsec-bias}

The anomalous diffusion that is biased by a constant external
force $F$ is a well established phenomenon found in many different
systems. For the biased CTRW the fractional rates
(\ref{f-rate1})-(\ref{f-rate2}) become site-independent, $f_i
\equiv f$ and $g_i \equiv g$, as $U_{i \pm 1}-U_i = \pm F \Delta
x$. From the fractional master equation~(\ref{FME2}) one finds
then the solutions for the mean particle position and for the mean
square displacement \cite{metzler1998},
\begin{eqnarray}
\langle x(t)\rangle &=& \langle x(0)\rangle + \frac{\Delta x (f -
g)} {\Gamma(\alpha + 1)} \, t^{\alpha} \, , \label{meanX} \\
\langle \delta x^2(t)\rangle &=& \langle \delta x^2(0)\rangle +
\frac{ \Delta x ^2 (f + g)}{\Gamma(\alpha + 1)} \, t^{\alpha}
\nonumber \\
&+& \left [\frac{2}{\Gamma(2 \alpha + 1)} -
\frac{1}{\Gamma^2(\alpha + 1)} \right] \Delta x^2 (f - g)^2 \,
t^{2 \alpha} \, . \label{meanX2}
\end{eqnarray}

The solutions of the corresponding fractional Fokker-Planck
equation are in the same form of the ones for the fractional
master equation; i.e.,
\begin{eqnarray}
\langle x(t)\rangle &=& \langle x(0)\rangle + \frac{F} {\eta_\alpha } \,
\frac{t^{\alpha}}{\Gamma(\alpha + 1)} \, , \label{FPX} \\
\langle \delta x^2(t)\rangle &=& \langle \delta x^2(0)\rangle +
 2 \kappa_\alpha  \, \frac{ t^{\alpha} }{\Gamma(\alpha + 1)}
\nonumber \\
&+& \frac{F^2}{\eta_\alpha^2} \, \left [\frac{2}{\Gamma(2 \alpha +
1)} - \frac{1}{\Gamma^2(\alpha + 1)} \right] \, t^{2 \alpha} \, .
\label{FPX2}
\end{eqnarray}
The comparison of the solutions (\ref{meanX}) to (\ref{FPX}) and
(\ref{meanX2}) to (\ref{FPX2}) gives,
\begin{equation} \label{currdiff}
\Delta x (f-g) = F/\eta_\alpha  \quad \mathrm{and} \quad \Delta
x^2 (f+g)/2 = \kappa_\alpha \, .
\end{equation}
The latter equations define the anomalous current and the anomalous diffusion
coefficient through the fractional rates $f$ and $g$, and are of
the same form as the corresponding relations for the normal
Brownian diffusion, determined through the corresponding escape rates.

If at a given temperature $T$ the system is close to thermal
equilibrium, the mean square displacement in the absence of an
external force and the average displacement induced by a bias $F
\ne 0$ are related through the {\it generalized Einstein relation}
\cite{bouchaud1990R, haus87, BarkaiFleurov1998},
\begin{equation} \label{einstein}
\left. \langle \delta x^2(t) \rangle \right|_{F=0} = \frac{ 2 }{
\beta F } \, \left. \langle x(t) - x(0) \rangle \right|_F \, .
\end{equation}
Note that equation~(\ref{einstein}) is strictly valid only in the linear
response regime, which is approached when when $F \to 0$. It then leads to
the generalized fluctuation-dissipation theorem
\begin{equation} \label{GFD}
\kappa_\alpha =
(\beta \eta_\alpha)^{-1} \, .
\end{equation}


\subsection{CTRW in a washboard potential} \label{subsec-wash}

Solving in the stationary limit the fractional Fokker-Planck
equation (\ref{FFPE}) for a biased periodic potential
$U(x)=U_0(x)- Fx$, one finds for the mean particle position
\cite{letter},
\begin{eqnarray} \label{x_per}
\langle x(t) \rangle = \langle x(0) \rangle +
\frac{v_{\alpha}(F)}{\Gamma( \alpha + 1)} \,  t^{\alpha} \, .
\end{eqnarray}
The anomalous current $v_\alpha (F)$ in the washboard potential is
then given by a generalized Stratonovich formula, put forward in
reference~\cite{letter}, i.e.,
\begin{equation} \label{stratSUB}
v_{\alpha}(F) = \frac{ \kappa_{\alpha} L \, [1 - \exp(-\beta F
L)]}{\int_{0}^L \mathrm{d} x \int_{x}^{x+L} \mathrm{d} y \,
\exp(-\beta[U(x) - U(y)])} \, .
\end{equation}

In analogy to the case with normal Brownian motion, in order to study the fractional
diffusion in a periodic or washboard potential, it would seem
natural to choose the lattice period $\Delta x$ to be equal to the
space period $L$ and the sites to be centered at minima, as
illustrated in figure~\ref{model}(b). In this case the fractional
rates, that we mark for such a lattice with $f_j$ and $g_j$, are
independent of the site $j$, $f_j \equiv \hat{f}$ and $g_j \equiv
\hat{g}$. It was proved in reference~\cite{letter} that
considering the CTRW in the lattice $\{ x_j = j L \}$, the
asymptotic solution ($t \to \infty $) for the mean square
displacement in a tilted periodic potential is of the form as the
solution (\ref{meanX2}) for the biased CTRW, as the
equation~(\ref{x_per}) is of the same form as
equation~(\ref{meanX}). The fractional rates $\hat{f}$
and $\hat{g}$, however, are no longer given by
equations~(\ref{f-rate1})-(\ref{f-rate2})  and the model
does not provide their explicit dependence on the potential.

Whereas in the washboard potential $\Delta x(\hat{f}-\hat{g})$ is
equal to the generalized Stratonovich current $v_\alpha$, one
could expect that $\Delta x^2 (\hat{f}+\hat{g})/2$ follows a
generalized formula for the effective diffusion coefficient in a
tilted periodic potential \cite{reimann01a,reimann01b}, in
correspondence to equations~(\ref{currdiff}) because the problem can be
mapped onto the case with a constant bias. However, in the long time
limit the ballistic term $\propto t^{2\alpha}$ prevails over the
term proportional to $t^\alpha$ and the effect of the latter one
is negligible: The ratio between the mean square displacement and
squared average coordinate depends in the asymptotic limit only on
the fractional exponent $\alpha$ and obeys the same result as  for the biased
CTRW \cite{ScherMontroll,letter}. The term proportional to
$t^\alpha$ becomes relevant for $t \to \infty$ only in the limit
$\alpha \to 1$, leading to the normal diffusive behavior, or for
$F \to 0$ as $\hat{f}-\hat{g} \to 0$, i.e., for a periodic
potential.


\section{Fractional diffusion in a periodic potential } \label{sec-result}

\subsection{The mean square displacement } \label{sec-meansq}

\begin{figure}[b]
\centering
\includegraphics[width=7.5cm]{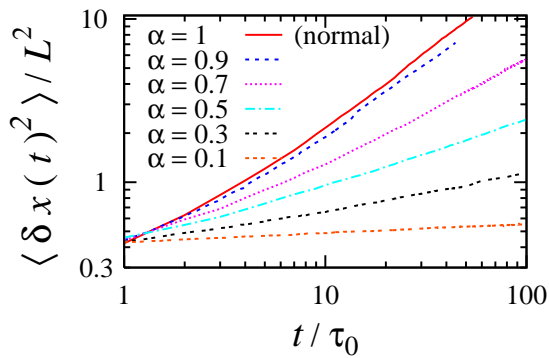}
\caption{(color online) The mean square displacement computed numerically for
the CTRW in the cosine-potential $U_0(x) = A\cos(2\pi x/L)$
for different values of the
fractional exponent $\alpha$. The re-scaled temperature is set at
$k_\mathrm{B}T/A=0.5$. } \label{meansq}
\end{figure}

In this section we present our results for fractional diffusion in spatially varying,
periodic potentials. We start from the expression of the mean
square displacement for the particle in the periodic potential
$U_0(x)$. For zero tilting the fractional rates $\hat{f}$ and
$\hat{g}$ become equal and the ballistic term occurring in
equation~(\ref{meanX2}), reformulated for a washboard potential, thus
disappears. Therefore, the asymptotic mean square displacement now
reads,
\begin{equation}
\langle \delta x^2(t)\rangle = \langle \delta x^2(0)\rangle +
\Delta x ^2 (\hat{f} + \hat{g}) \, \frac{t^{\alpha}}{\Gamma(\alpha
+ 1)} \, . \label{meanX20}
\end{equation}
This equation is confirmed by the numerical results, depicted in
figure~\ref{meansq} for various values of the fractional exponent
$\alpha$, which is equal to the slope of the numerically
evaluated curves on the logarithmic scale.

Correspondingly, the asymptotic solution of the fractional
Fokker-Planck equation for the periodic potential $U_0(x)$ can be
written in the form analogous to the case of the fractional
diffusion in the absence of force, i.e.,
\begin{equation}
\langle \delta x^2(t)\rangle = \langle \delta x^2(0)\rangle + 2 \,
\kappa_\alpha^\mathrm{(eff)} \frac{t^{\alpha}}{\Gamma(\alpha + 1)}
\, , \label{FPX20}
\end{equation}
This equation defines  the {\it effective fractional diffusion
coefficient} $\kappa_\alpha^\mathrm{(eff)}$,
\begin{equation} \label{kappa-def}
\kappa _\alpha ^{(\mathrm{eff})} = \Gamma (\alpha +1) \lim _{t \to
\infty} \frac{\langle \delta x^2(t)\rangle - \langle \delta
x^2(0)\rangle }{2t^\alpha} \, .
\end{equation}
Also the explicit expression for the fractional rates to hop from
one minimum to one of the neighboring minima follows from
equations~(\ref{meanX20}) and (\ref{FPX20}), as for $\hat{f}
\equiv \hat{g}$
\begin{equation}
\kappa _\alpha ^{(\mathrm{eff})} = \frac{\Delta x ^2 (\hat{f} +
\hat{g})}{2} \equiv \Delta x ^2 \hat{f} \equiv \Delta x ^2 \hat{g}
\, .
\end{equation}
%


\subsection{Effective fractional diffusion coefficient} \label{sec-coefficient}

Next, a useful analytical expression for $\kappa _\alpha ^{(\mathrm{eff})}$ in
a periodic potential can be derived from a generalized
Einstein relation. Equation~(\ref{einstein}) is valid in the
linear response regime also for any periodic potential $U_0(x)$,
as the problem of fractional diffusion in a periodic potential can
be mapped onto the  force free case~\cite{letter}. In doing so we
can write
\begin{equation}
\kappa_\alpha^{(\mathrm{eff})}= \frac{1}{\beta} \lim _{F \to 0}
\frac{ v_\alpha(F)}{F} = \left . \frac{1}{\beta} \frac{ \mathrm{d}
v_\alpha(F)}{\mathrm{d} F} \right |_{F=0} \, ,
\end{equation}
where $v_\alpha (F)$ is given by the generalized Stratonovich
formula derived in (\ref{stratSUB}). As a central result we thus
obtain the following closed, exact analytical expression for the
effective fractional diffusion coefficient, reading
\begin{equation} \label{FLJ}
\kappa^{(\mathrm{eff})}_\alpha = \frac{ \, \kappa_\alpha}{ L^{-2}
\int_0^L \mathrm{d}x \, \exp\left[\beta U_0(x)\right] \int_0^L
\mathrm{d}y \, \exp\left[-\beta U_0(y)\right] } \, .
\end{equation}
This expression is valid for an arbitrary shaped, unbiased
periodic potential $U_0(x)$. It reduces for $\alpha = 1$ to the
corresponding formula for the normal diffusion in a periodic
potential, first derived by Lifson and Jackson in
reference~\cite{lifson1962} and independently  again in
references~\cite{festa1978, weaver1979}. Our new result therefore
provides the generalization for fractional diffusion processes
which are anomalous.

\begin{figure}[b]
\centering
\includegraphics[width=7.5cm]{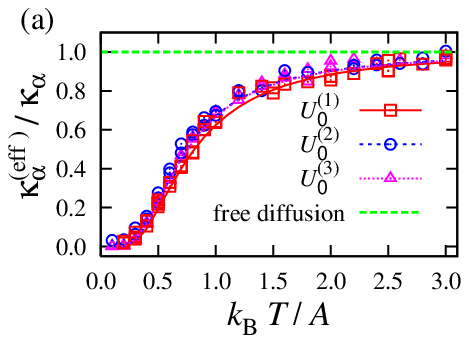}
\includegraphics[width=7.5cm]{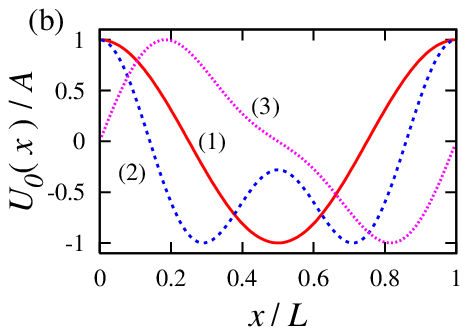}
\caption{(color online) (a) Effective anomalous diffusion coefficient
$\kappa_\alpha^\mathrm{(eff)}$ in a periodic potential {\it
versus} the re-scaled temperature $k_\mathrm{B}T/A$. The quantity
$\kappa_\alpha^{\mathrm{(eff)}}$ is re-scaled by the corresponding
free fractional diffusion coefficient $\kappa_\alpha$. The
theoretical curves obtained from equation~(\ref{FLJ}) (lines) are compared
to the numerical results (symbols). The different periodic
potentials used are given by
equations~(\ref{pot-cos})-(\ref{pot-rat}). For each potential and at
 given temperature the numerical points are computed for some values
of $\alpha$ within the interval $\alpha \in [0.1, 0.9]$.
(b) A comparison among the different periodic potentials
used for the numerics, see in (\ref{pot-cos})-(\ref{pot-rat}):
(1) the cosine potential $U_0^{(1)}(x)$; (2) the double
hump potential $U_0^{(2)}(x)-(2 a_1 -1)$; (3) the ratchet
potential $U_0^{(3)}(x)$. } \label{fig-FLJ}
\end{figure}

The behavior of equation~(\ref{FLJ}) \textit{versus} re-scaled
temperature $k_\mathrm{B} T / A$ is illustrated in
figure~\ref{fig-FLJ}(a) for the following
periodic potentials, depicted with figure~\ref{fig-FLJ}(b): \\
(i) a cosine potential,
\begin{equation} \label{pot-cos}
U_0^{(1)}(x) = A \cos (2 \pi x/L) \, ,
\end{equation}
(ii) a double hump potential,
\begin{equation} \label{pot-dh}
U_0^{(2)}(x) = A a_1 [ \cos (2 \pi x/L) + \cos (4 \pi x/L) ] \, ,
\end{equation}
with the coefficient $a_1 = 16/25$, and \\
(iii) a ratchet potential,
\begin{equation} \label{pot-rat}
U_0^{(3)}(x) = A [a_2 \sin (2 \pi x/L) + a_3 \sin (4 \pi x/L) ] \,
,
\end{equation}
with $a_2 = 85/(21 \sqrt{21})$, $a_3 = 25 /(21 \sqrt{21})$. The
coefficients $a_1$, $a_2$, $a_3$ are chosen such that the
potentials (\ref{pot-cos})-(\ref{pot-rat}) have the same amplitude
$A$. The theoretical curves are confirmed by numerical results,
depicted in figure~\ref{fig-FLJ} with symbols. The anomalous
diffusion coefficient is computed as defined by
equation~(\ref{kappa-def}). As the ratio
$\kappa_\alpha^{(\mathrm{eff})}/\kappa_\alpha < 1$, one can
conclude that, analogously to the normal case, the effect of any
one-dimensional non-biased periodic field is to {\it suppress} the
macroscopic anomalous diffusion coefficient compared to the value
in the absence of force \cite{lifson1962}. A possible  enhancement
may be expected in presence of time-dependent, periodic landscape
modulations as demonstrated for normal diffusion in
references~\cite{ENH1,ENH2,ENH3}. Furthermore, it is to be noticed
that the ratio $\kappa_\alpha^{(\mathrm{eff})}/\kappa_\alpha $
does {\it not} depend on the fractional exponent $\alpha$ and
moreover, the shape of the periodic potential $U_0(x)$ has only a
small influence, as one can see by comparing the theoretical
curves in figure~\ref{fig-FLJ} (a)  (note also
references~\cite{heinsalu2004a, heinsalu2005a}).


\subsection{Probability density: Anomalous versus normal}
\label{sec-prob}

\begin{figure}[b]
\centering
\includegraphics[width=15.0cm]{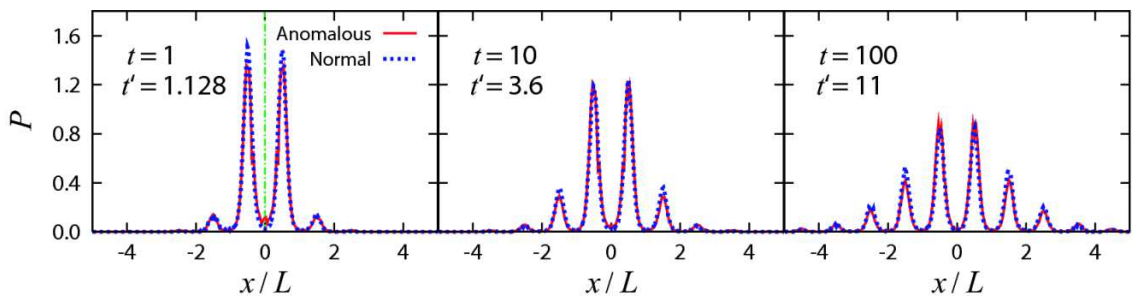}
\caption{(color online) The time evolutions of the probability densities
characterizing the anomalous and normal diffusion processes in the periodic
cosine potential (\ref{pot-cos}).
The re-scaled temperature is $k_\mathrm{B} T / A = 0.5$
and the fractional exponent is $\alpha = 0.5$. The anomalous
probability density $P(x,t)$ cannot be distinguished from that of
the normal case, $P(x,t')$, once the time has been re-scaled according
to equation~(\ref{time}). Similar results are obtained for other
values of $\alpha \in (0, 1)$ (not depicted). } \label{dist-F0}
\end{figure}
\begin{figure}[t]
\centering
\includegraphics[width=7.5cm]{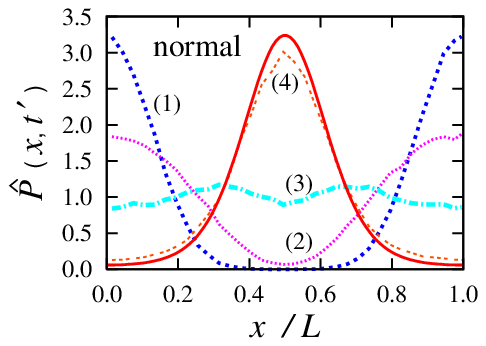}
\includegraphics[width=7.5cm]{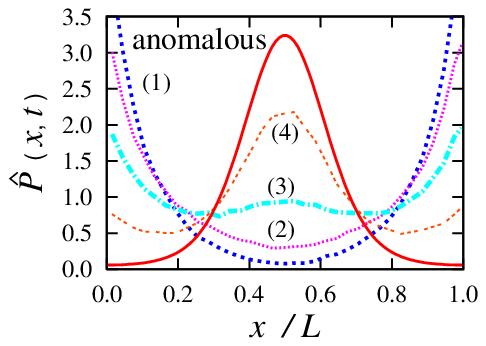}
\caption{(color online) The different small-time evolutions of the normal (left) and
  anomalous (right) reduced probability densities
$\hat{P}(x, t)$ and $\hat{P}(x, t')$ defined by equation (\ref{P1}),
in the cosine potential (\ref{pot-cos}).
Curve labels (1), (2), (3) and (4) represent increasing values
of re-scaled time $t' = 0.01, 0.02, 0.04, 0.11$ for the normal case and
of time $t$ for the anomalous case, related to $t'$ through equation~(\ref{time}).
The solid line represents the theoretical stationary solution.
The re-scaled temperature is $k_\mathrm{B}T / A =
0.5$ and $\alpha = 0.5$ for the anomalous process, as in
figure~\ref{dist-F0}. } \label{reduced}
\end{figure}

In the previous section it was demonstrated that the effective fractional
diffusion coefficient in a periodic potential is of the same form as the
the Lifson-Jackson formula for  normal diffusion. This
represents a further element of the formal analogy between fractional
and normal diffusion, besides e.g. the validity of the generalized
Stratonovich formula (\ref{stratSUB}) ~\cite{letter} and the fact that the
stationary reduced probability density is the same for both
cases \cite{heinsalu2006b}. Here, we present  additional results
which support and corroborate this formal analogy further. We notice that in the absence
of a bias, all the odd moments of the probability density are identically zero
both for normal and fractional diffusion. As for the
second moment, upon introducing the re-scaled time,
\begin{equation}\label{time}
t' = \frac{(t/\tau_0)^\alpha}{\Gamma(1+ \alpha)} \, ,
\end{equation}
it follows from equation~(\ref{FPX20}) that the mean square
displacement (in units of $L^2$) formally coincides
with that of the normal diffusion case,
$[\langle \delta x^2 (t') \rangle - \langle \delta x^2 (0)
\rangle] / L^2 = 2 \, T' \, t'$,
independently of the fractional exponent $\alpha$,
wherein $T' = k_\mathrm{B} T / A$, with $A$  the potential amplitude,
is the re-scaled temperature. The study of the time evolution of the
probability density is illustrated with the example in figure~\ref{dist-F0}
choosing the times $t$ for the anomalous diffusion process 
and the corresponding times $t'$ for normal diffusion, so that they satisfy
equation (\ref{time}): The probability
densities for anomalous diffusion (continuous lines) and normal diffusion (dashed
lines) processes are barely distinguishable from each other for sufficiently
long evolution times.

In clear contrast, however, appreciable differences between the normal diffusion
coordinate density $P(x,t')$ and the anomalous coordinate density $P(x,t)$
emerge for small times. This is best detectable by comparing
the reduced probability density, mapped onto a single
spatial period,
\begin{equation} \label{P1}
\hat{P}(x, t) = \sum_n P(nL+x, t) \, , \quad n \in \mathbb{Z}  \,
,
\end{equation}
as done in figure~\ref{reduced}.
In the normal case (figure~\ref{reduced} left) the two initial maxima
at $x=0$ and $x/L=1$, due to
the initial conditions $\hat{P}(x,0) = \delta(x)$,
move toward the center and finally merge into the asymptotic stationary
density (solid line) $\hat{P}_\mathrm{st}(x) =
\mathcal{N}^{-1}\exp[- \beta U_o(x)]$, where $\mathcal{N} = \int_0^1
dx' \exp[- \beta U_o(x')]$ is a normalization factor and
$U_0(x) = A\cos(2\pi x/L)$.
On the other hand, in the anomalous case the two initial maxima
gradually disappear, while a new peak grows at $x/L = 0.5$ and evolves into
the stationary density $\hat{P}_\mathrm{st}(x)$.
\begin{figure}[t]
\centering
\includegraphics[width=15.0cm]{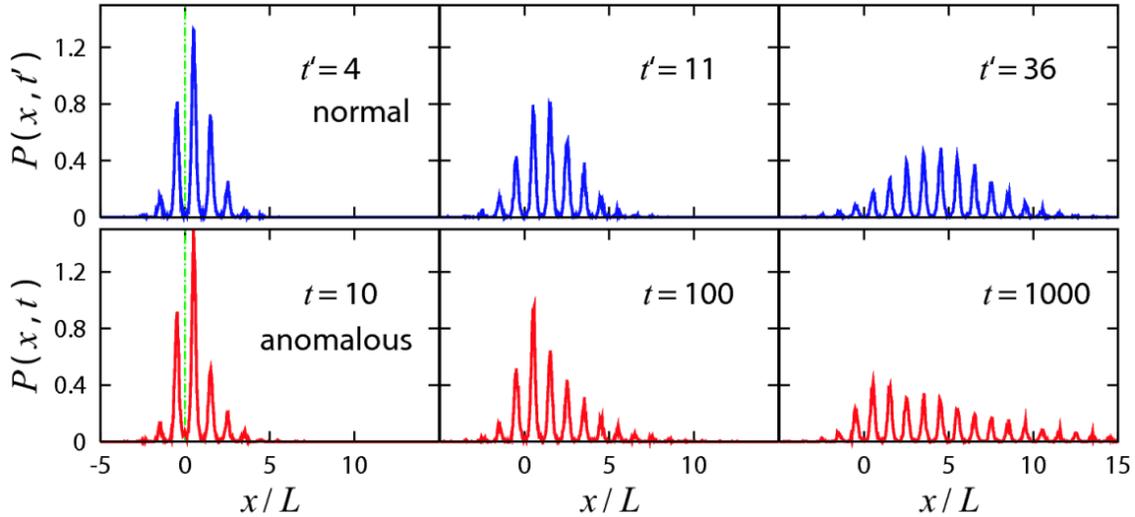}
\caption{(color online) The time evolutions of the probability densities
characterizing the normal (above) and anomalous (below) diffusion
processes in a tilted cosine potential $U(x)= A\cos(2 \pi x/L)-Fx$.
The re-scaled temperature is $k_\mathrm{B} T / A = 0.5$ and the fractional
exponent is $\alpha = 0.5$, as in figure~\ref{dist-F0}. The
tilting force is $F=0.1 \times F_\mathrm{cr}$, where $F_\mathrm{cr} = 2
\pi A/L$ is the re-scaled critical bias, corresponding to the
disappearance of potential minima. For sufficiently small times
the probability densities of the normal and anomalous processes
are very similar. However, at larger times (in the long time
limit) the maximum of the density for normal diffusion moves with the directed current.
In contrast, the mean square displacement of an ensemble of particles
undergoing fractional diffusion is dominated by the ballistic
contribution and the typical stretched spreading in the direction
of bias is observed, while leaving the maximum of the density near the origin. } \label{dist-F}
\end{figure}

Moreover, as soon as the process is biased by an external finite force, $F \ne
0$, a qualitative difference arises in the time evolutions of the
probability densities of the anomalous and the normal processes in the long
time limit as well, see also Ref.~ \cite{heinsalu2006b}. This is true even for small values of $F$ in
the linear response regime, as one can defer from  figure~\ref{dist-F}.
All this indicates a profound difference between a fractal diffusion dynamics that is  based on
the fractal Brownian motion introduced by Mandelbrot and van Ness \cite{mandelbrot1968}
and the fractional diffusion based on the CTRW \cite{MontrollWeiss}.
The time evolution of the density of an ensemble of particles
undergoing normal diffusion can be interpreted as a superposition
of a translational motion and a spreading of the initially
localized density. In this case one observes the global maximum of
the probability density moving in the direction of the external bias
(figure~\ref{dist-F}~(above)). Instead, in the anomalous case,
only a spreading of the initial density takes place, resulting in
a long tail in the direction of the bias. The global maximum
of the density remains close, however, to its initial position
(figure~\ref{dist-F}~(below)) \cite{heinsalu2006b}. This intriguing
behavior is related to the presence of a ballistic contribution
proportional to $t^{2\alpha}$ in the mean square displacement (see
equation~(\ref{meanX2})). We remark that for $\alpha$ close to one
and for small values of external bias $F$, at small times the term
$\propto t^\alpha$ can prevail the ballistic term. However, in the
long-time limit the ballistic term takes over and always dominates. The latter
remark may be relevant for experimental studies. It in addition also provides a
crucial test that allows one to distinguish between fractal and fractional Brownian motion
on a practical level.


\section{Conclusion}
With this work we investigated anomalous diffusion  whose dynamics is governed by
a fractional Fokker-Planck equation with a spatially varying, periodic potential.
As a main result we  derive a generalization of the celebrated Lifson-Jackson
result for normal diffusion \cite{lifson1962,festa1978, weaver1979}
to our case with anomalous fractional diffusion: It  relates the effective
fractional diffusion coefficient $\kappa^{(\mathrm{eff})}_\alpha$ in Eq.~(\ref{FLJ}) to
the bare fractional diffusion coefficient  in terms of two inverse
quadratures of the periodic potential only. As a consequence, we find that like
in the case with normal diffusion the effective anomalous diffusion becomes always
suppressed over the bare value. This result may find ample application in diverse areas where
anomalous diffusion occurs; typical examples are the case of superionic conductors
\cite{SIC} or for the Josephson junction dynamics \cite{HTB,Kautz}
when the role of disorder may change  the normal diffusion into anomalous one.

In addition, we contrasted the time evolution for normal
diffusion with anomalous, fractional diffusion. In doing so, we find that after
a proper re-scaling of time the corresponding asymptotic densities $P(x,t)$
for the coordinate $x$ match each other. Distinct differences occur, however,
at small evolution times. This time evolution of the densities drastically
changes upon the application of a finite bias $F$. Now, the long time  evolution
between normal diffusion and anomalous diffusion becomes markedly distinct as well: While the maximum
of the biased  normal diffusion moves with the normal, directed current, the anomalous
case is dominated by a ballistic spreading that leaves the maximum of the density around the origin.
Moreover, this characteristic difference can be put to work to differentiate between fractional and
fractal Brownian diffusion.


\ack

This work has been supported by the Estonian Science Foundation
through grant no. 6789 and by the Archimedes Foundation (EH), by
the DFG via the collaborative research center, SFB-486, project A10, and by the
Volks\-wagen Foundation, via project no. I/80424.


\end{document}